\documentclass[11pt]{article}
\usepackage{moriond,epsfig}

\def\be{\begin{equation}}
\def\ee{\end{equation}}
\def\bea{\begin{eqnarray}}
\def\eea{\end{eqnarray}}

\begin{document}
\vspace*{4cm}
\title{
Production of digluon and quark-antiquark dijets in central exclusive
processes
}

\author{Antoni Szczurek $^{1,2}$}

\address{$^{1}$ {\em Institute of Nuclear Physics PAN\\
PL-31-342 Cracow, Poland\\}
$^{2}$ {\em University of Rzesz\'ow\\
PL-35-959 Rzesz\'ow, Poland\\}
}

\maketitle

\abstracts{
We discuss exclusive central production of Higgs boson,
quark-antiquark and digluon dijets. Some differential distributions
are shown and disussed.
Irreducible leading-order $b \bar b $ background to Higgs production is 
calculated.
The signal-to-background ratio is shown and improvements are 
suggested by imposing cuts on $b$ ($\bar b$) transverse momenta
and rapidities. We disuss also gluonic dijet production. Here we use
rather reggeon-reggeon-gluon vertices. We discuss briefly also a new
mechanism of emission of gluons from different $t$-channel gluons (reggeons).
The latter contribution turned out to be rather small.
When gluons are missidentified as $b$ or $\bar b$ jets the latter
contribution constitutes a reducible but large contribution to
exclusive Standard Model Higgs boson.
}

\section{Introduction}

Since the cross section for exclusive Higgs boson production is rather
small, only $b \bar b$ final state can be used in practice to identify Higgs
boson. This means that a $b \bar b$ continuum background is of crucial
importance. We discuss this irreducible background here.

In our calculations we include exact matrix elements
and do full four-body calculations for all considered processes. 
The kinematically complete calculations allow to include any cut on 
kinematical variables which is very usefull in order to find
the Higgs boson signal.

We consider also exclusive gluonic dijet production. Such dijets has
been observed experimentally by the CDF collaboration at the Tevatron
\cite{CDF-dijets} and constitute a benchmark for further exclusive Higgs
boson studies.

\section{Formalism}

A modern approach to exclusive Higgs boson production was
proposed by Khoze, Martin and Ryskin \cite{KMR_Higgs}.
Here we discuss this approach for exclusive production of
quark-antiquark and digluon dijets.

\subsection{$p p \to p p q \bar q$}

\begin{figure}[!htp]
\begin{center}
\includegraphics[width=8cm]{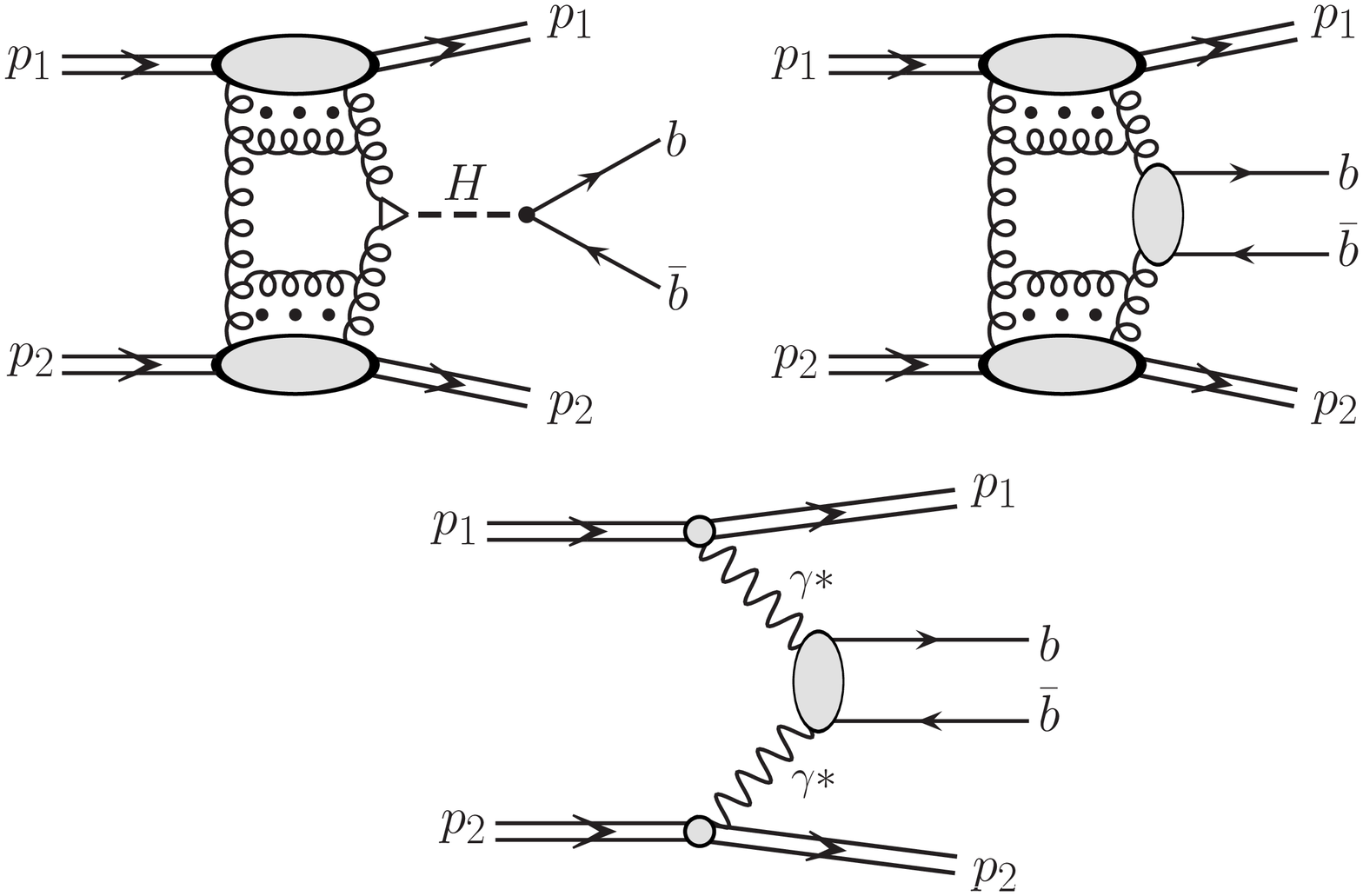}
\caption{Diagrams included in the calculation of $b \bar b$ jets.
}
\label{fig:new_diagrams}
\end{center}
\end{figure}

Let us concentrate on the simplest case of the production of $q\bar{q}$
pair in the color singlet state. We do not consider the $q \bar q g$
contribution as it is higher order compared to the one considered here.
In Refs.\cite{MPS10higgs,MPS11higgs} the mechanisms for 
$b \bar b$ production shown in Fig.\ref{fig:new_diagrams} have been considered.

We write the amplitude of the exclusive diffractive $q\bar{q}$ pair
production $pp\to p(q\bar{q})p$ in the color singlet state as \\
\begin{eqnarray}
{\cal M}_{\lambda_q\lambda_{\bar{q}}}^{p p \to p p q \bar
  q}(p'_1,p'_2,k_1,k_2)& = &
s\cdot\pi^2\frac12\frac{\delta_{c_1c_2}}{N_c^2-1}\, 
\Im \int d^2
q_{0,t} \; V_{\lambda_q\lambda_{\bar{q}}}^{c_1c_2}(q_1, q_2, k_1, k_2) 
\; \nonumber \\
&& \frac{f^{\mathrm{off}}_{g,1}(x_1,x_1',q_{0,t}^2,
q_{1,t}^2,t_1) \; f^{\mathrm{off}}_{g,2}(x_2,x_2',q_{0,t}^2,q_{2,t}^2,t_2)}
{q_{0,t}^2\,q_{1,t}^2\, q_{2,t}^2} \; ,
\label{amplitude}
\end{eqnarray}
where $\lambda_q,\,\lambda_{\bar{q}}$ are helicities of heavy $q$
and $\bar{q}$, respectively. Above $f_1^{\mathrm{off}}$ and
$f_2^{\mathrm{off}}$ are the off-diagonal unintegrated gluon
distributions in nucleon 1 and 2, respectively. 
The longitudinal momentum fractions of active gluons
are calculated based on kinematical variables (transverse masses 
and rapidities) of outgoing quark
and antiquark.
The bare amplitude above is subjected to absorption corrections.
The absorption corrections are taken here in a simple multiplicative form.

The color singlet $q\bar{q}$ pair production amplitude can be written as
\cite{MPS11higgs}
\begin{equation}
V_{\lambda_q\lambda_{\bar{q}}}^{c_1c_2}(q_1,q_2,k_1,k_2)\equiv
n^+_{\mu}n^-_{\nu}V_{\lambda_q\lambda_{\bar{q}}}^{c_1c_2,\,\mu\nu}
(q_1,q_2,k_1,k_2) \; .
\nonumber
\end{equation}
The tensorial part of the amplitude reads:
\begin{equation}
\begin{array}{lll}
V_{\lambda_q\lambda_{\bar{q}}}^{\mu\nu}(q_1, q_2, k_1, k_2)
= g_s^2 \,\bar{u}_{\lambda_q}(k_1)
\biggl(\gamma^{\nu}\frac{\hat{q}_{1}-\hat{k}_{1}-m}
{(q_1-k_1)^2-m^2}\gamma^{\mu}-\gamma^{\mu}\frac{\hat{q}_{1}
-\hat{k}_{2}+m}{(q_1-k_2)^2-m^2}\gamma^{\nu}\biggr)v_{\lambda_{\bar{q}}}(k_2).
\end{array}
\end{equation}
The coupling constants $g_s^2 \to g_s(\mu_{r,1}^2)
g_s(\mu_{r,2}^2)$. In the present calculation we take the
renormalization scale to be $\mu_{r,1}^2=\mu_{r,2}^2=M_{q \bar q}^2$. 
The exact matrix element is calculated numerically. Analytical formulae 
are shown explicitly in \cite{MPS11higgs}.

\subsection{$p p \to p p g g$}

The exclusive digluon production has been discussed before in
\cite{CDHI2009,DKRS2011,MPS11_gluons}.


\begin{figure} [!thb]
\begin{center}
\includegraphics[width=8cm]{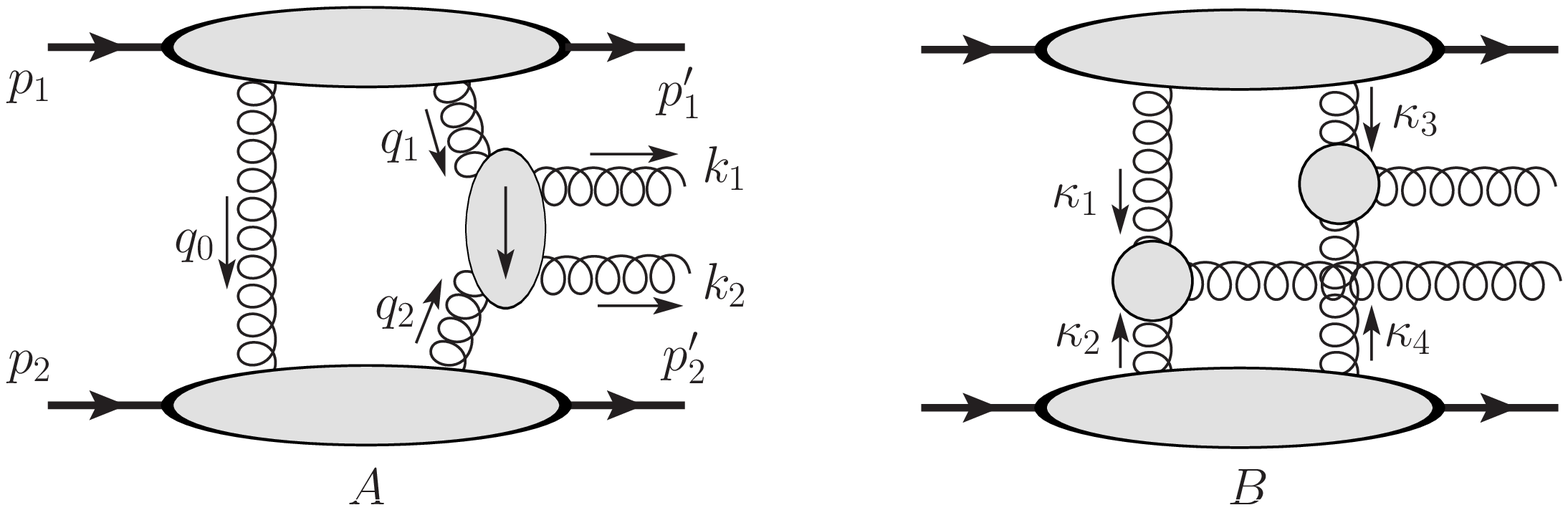}
\end{center}
\caption{Mechanisms of exclusive gluonic dijet production.}
\label{fig:gg_CEP}
\end{figure}


The mechanisms of digluon production are shown in Fig.\ref{fig:gg_CEP}.
The matrix element for diagrams Fig.~\ref{fig:gg_CEP} (A) and
(B) to the diffractive amplitude ${\cal M}^{gg} = {\cal M}^{A} +
{\cal M}^{B}$ for the central exclusive $gg$ (with external color
indices $a$ and $b$) dijet production $pp\to p(gg)p$ can be written as
\cite{MPS11_gluons}
\begin{eqnarray}\nonumber
{\cal M}^{A}_{ab}(\lambda_1,\lambda_2)&=&is\,{\cal
A}\,\frac{\delta_{ab}}{N_c^2-1}\int d^2{\bf
q}_0\frac{f^{\mathrm{off}}_g(q_0,q_1)f^{\mathrm{off}}_g(q_0,q_2)\cdot
\epsilon^*_{\mu}(\lambda_1)\epsilon^*_{\nu}(\lambda_2)}{{\bf
q}_0^2{\bf q}_1^2{\bf q}_2^2}\,\times\\
&&\left[\frac{C_1^{\mu}(q_1,r_1)C_2^{\nu}(r_1,-q_2)}{{\bf r}_1^2}
+\frac{C_1^{\mu}(q_1,r_2)C_2^{\nu}(r_2,-q_2)}{{\bf r}_2^2}\right]\,,\label{ampl-a+b}\\
{\cal M}^{B}_{ab}(\lambda_1,\lambda_2)&=&-is\,{\cal
A}\,\frac{\delta_{ab}}{N_c^2-1}\int d^2{\bf
\kappa}_1\frac{f^{\mathrm{off}}_g(\kappa_1,\kappa_3)f^{\mathrm{off}}_g(\kappa_2,\kappa_4)\cdot
\epsilon^*_{\mu}(\lambda_1)\epsilon^*_{\nu}(\lambda_2)}
{{\bf \kappa}_1^2{\bf \kappa}_2^2{\bf \kappa}_3^2{\bf \kappa}_4^2}\,
\times
\nonumber \\
&&C_1^{\mu}(\kappa_1,-\kappa_2) C_2^{\nu}(\kappa_3,-\kappa_4),
\label{amplitude_gg}
\end{eqnarray}
where ${\cal A}=2\pi^2g_s^2/C_F$, the minus sign in ${\cal M}^{B}$
comes from the difference in colour factors,
$\epsilon^*_{\mu}(\lambda_1)$ and $\epsilon^*_{\nu}(\lambda_2)$ are
the polarisation vectors of the final state gluons
with helicities $\lambda_1,\,\lambda_2$ and momenta $p_3,\,p_4$,
respectively, $f^{\mathrm{off}}_g(v_1,v_2)$ is the off-diagonal
UGDF, which is dependent on longitudinal and transverse components
of both gluons with 4-momenta $v_1$ and $v_2$, emitted from a single
proton line, and
\[r_2=q_1-p_4\,,\quad\kappa_2=-(\kappa_1-p_4)\,,\quad\kappa_4=-(\kappa_3-p_3)\,.\]

\subsection{Off-diagonal unintegrated gluon 
distributions}

The off-diagonal parton distributions (i=1,2) are calculated as
\begin{equation}
\begin{array}{lll}
f_i^{\mathrm{KMR}}(x_i,Q_{i,t}^2,\mu^2,t_i)  =  R_g
\frac{d[g(x_i,k_t^2)S_{1/2}(k_{t}^2,\mu^2)]}{d \log k_t^2} |_{k_t^2
= Q_{it}^2} \;
F(t_i) \; ,
\end{array}
\label{KMR-off-diagonal-UGDFs}
\end{equation}
where $S_{1/2}(q_t^2, \mu^2)$ is a Sudakov-like form factor relevant
for the case under consideration. 
It is reasonable to take a factorization scale as: $\mu_1^2 =
\mu_2^2 = M_{q \bar q}^2, M_{gg}^2$.

The factor $R_g$ here cannot be calculated from first principles
but can be estimated in the case of off-diagonal collinear PDFs
when $x' \ll x$ and $x g = x^{-\lambda}(1-x)^n$.
Typically $R_g \sim$ 1.3 -- 1.4 at the Tevatron energy. 
The off-diagonal form factors are parametrized here as 
$F(t) = \exp \left( B_{\mathrm{off}} t \right)$.
In practical calculations we take $B_{\mathrm{off}}$ = 2 GeV$^{-2}$.
In evaluating $f_1$ and $f_2$ needed for calculating the amplitudes
(\ref{amplitude},\ref{amplitude_gg}) we use different collinear distributions.

In the case of diagram B for digluon production we have to
notice that we are in the ERBL region \cite{MPS11_gluons}.
The details how to treat then unintegrated gluon distributions
was discussed in detail in \cite{MPS11_gluons}.

\section{Results}

\subsection{Production of $b \bar b$ jets}

The Higgs boson differential cross sections are calculated assuming
a three-body process $p p \to p H p$.
Assuming full coverage for outgoing protons
we construct two-dimensional distributions 
$d \sigma / dy d^2 p_t$ in Higgs rapidity and transverse momentum. 
The distribution is used then in a simple Monte Carlo code 
which includes the Higgs boson decay into the $b {\bar b}$ channel. 
It is checked whether $b$ and $\bar b$ enter into the central detector. 

\begin{figure}[!htp]
\begin{center}
\includegraphics[width=6cm]{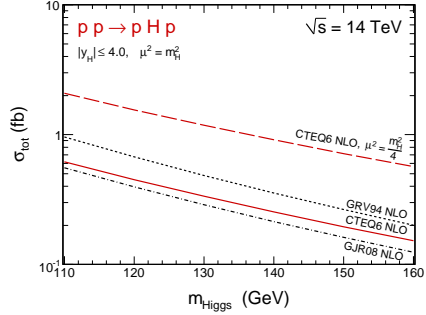}
\caption{Total cross section as a function of Higgs boson mass.
}
\label{fig:sig_MH}
\end{center}
\end{figure}

In Fig.\ref{fig:sig_MH} we show the Higgs boson production cross section
as a function of Higgs boson mass for different collinear gluon distributions.
The cross section is rather small.

\begin{figure}
\begin{center}
\includegraphics[width=6.0cm]{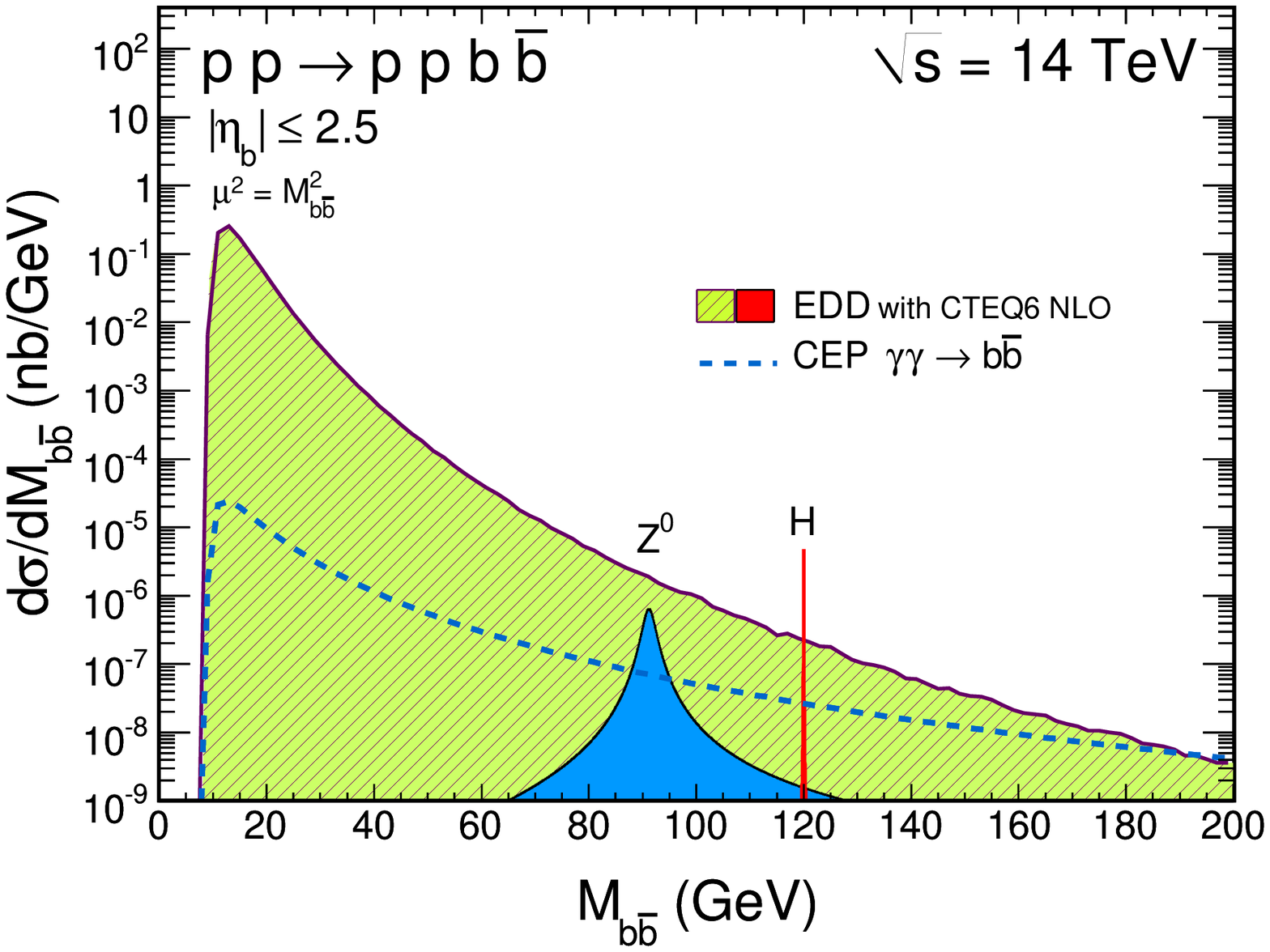}
\includegraphics[width=6.0cm]{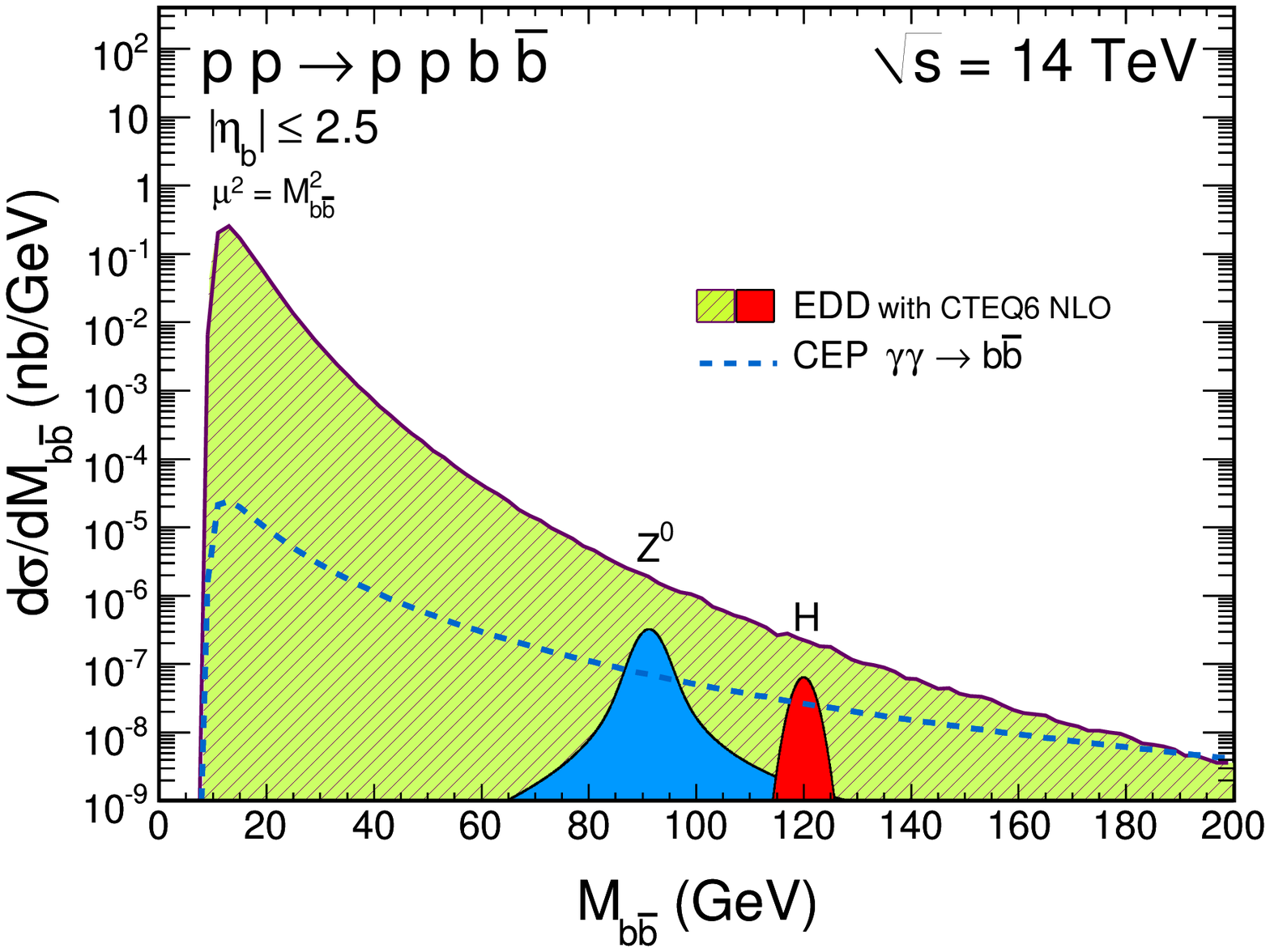}
\end{center}
\caption{The $b \bar b$ invariant mass distribution for $\sqrt{s}$ =
14 TeV and for $b$ and $\bar b$ jets for $-2.5
< \eta < 2.5$ corresponding to the ATLAS detector. 
The left panel shows purely theoretical predictions, while the right
panel includes experimental effects due to experimental uncertainty
in missing mass measurement.} 
\label{fig:dsigma_dMbb_fully}
\end{figure}

In the left panel of Fig.\ref{fig:dsigma_dMbb_fully} we show the 
central diffractive contribution for CTEQ6 \cite{CTEQ} 
collinear gluon distribution and the contribution from 
the decay of the Higgs boson including natural decay width, see
the sharp peak at $M_{b \bar b}$ = 120 GeV.
The phase space integrated cross section for the Higgs production,
including absorption effects with gap survival probability $S_G = 0.03$ 
is less than 1 fb.
The result shown in Fig.\ref{fig:dsigma_dMbb_fully} includes 
branching fraction for BR($H \to b \bar b) \approx$ 0.8 
and the rapidity restrictions. The much broader 
Breit-Wigner type peak to the left of the Higgs signal corresponds to 
the exclusive production of the $Z^0$ boson with the cross section 
calculated as in Ref.~\cite{CSS09}.
The branching fraction BR($Z^0 \to b \bar b) \approx$ 0.15 
has been included in addition. In contrast to the
Higgs case the absorption effects for the $Z^0$ production are much
smaller \cite{CSS09}. The sharp peak
corresponding to the Higgs boson clearly sticks above the
background. 

\begin{figure}[!h]
\begin{center}
\includegraphics[width=5cm]{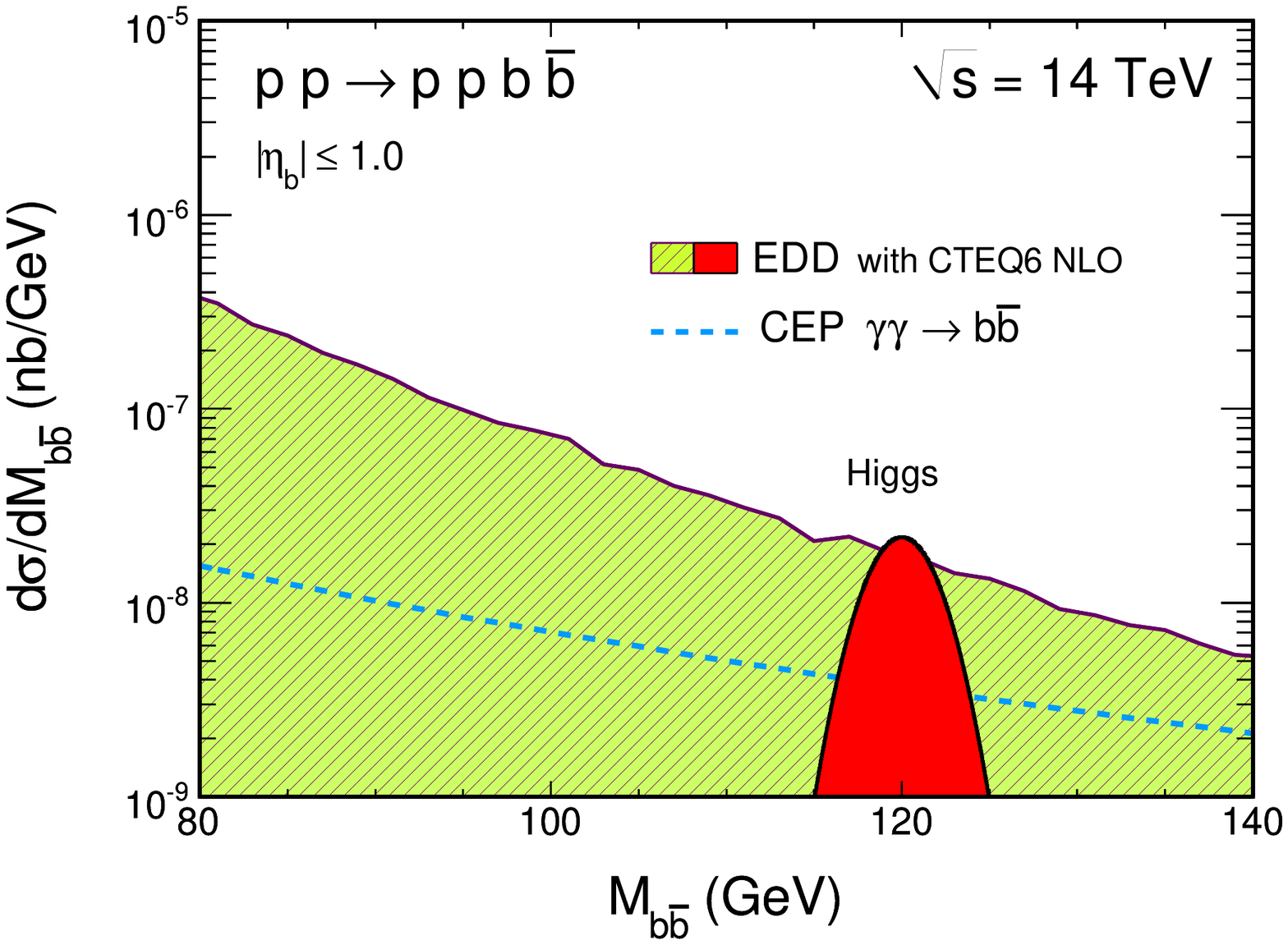}
\includegraphics[width=5cm]{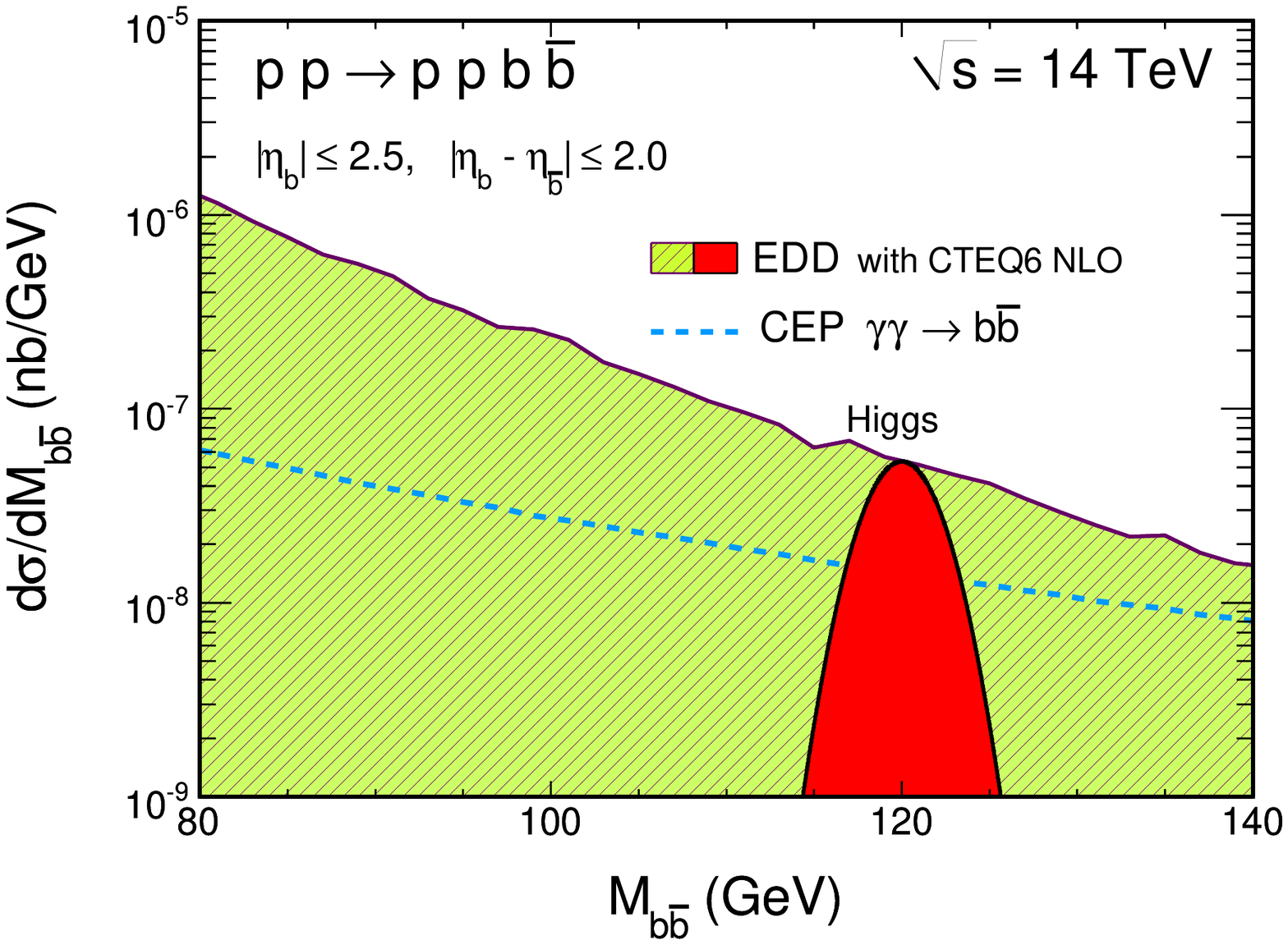} \\
\includegraphics[width=5cm]{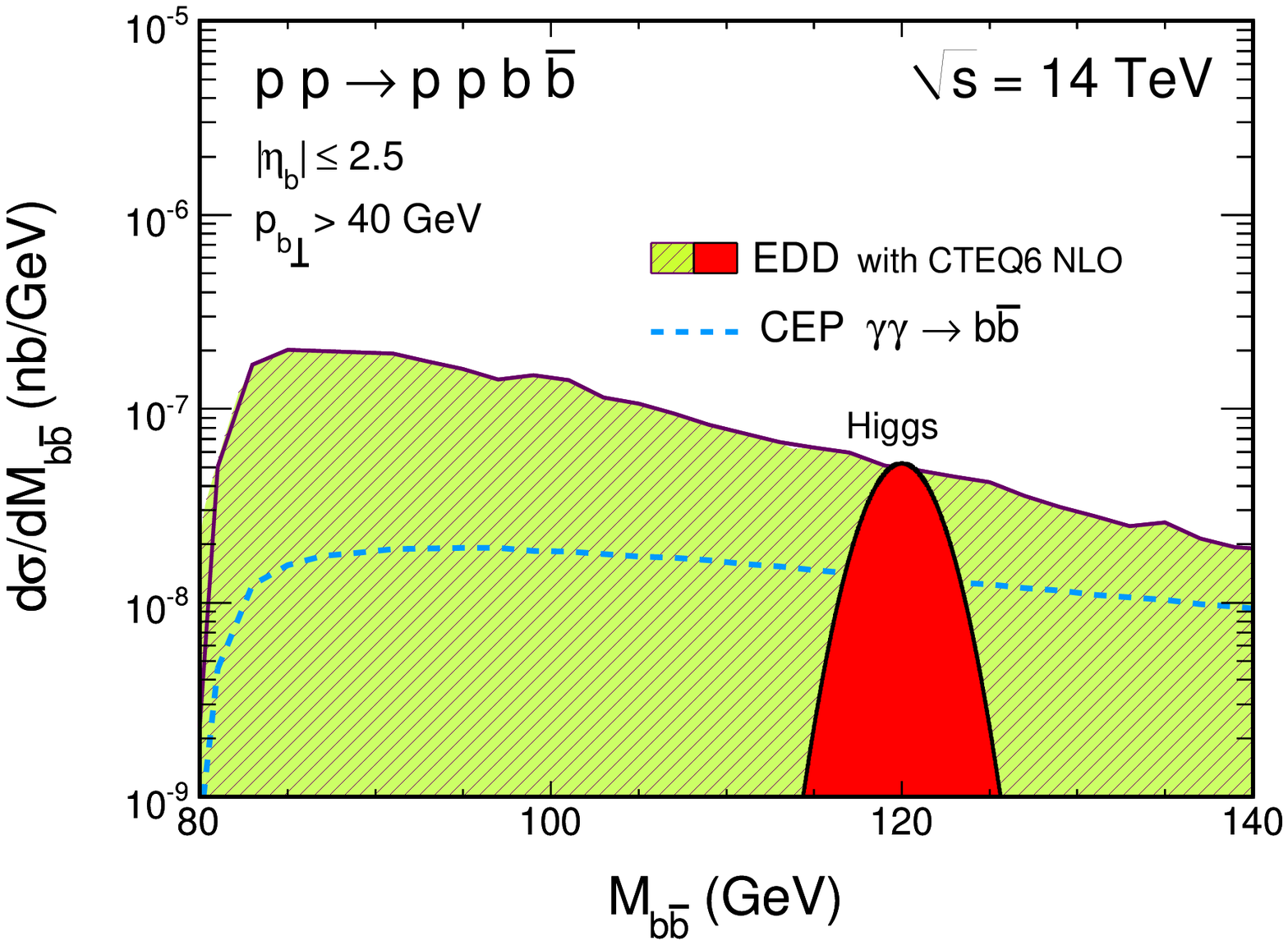}
\includegraphics[width=5cm]{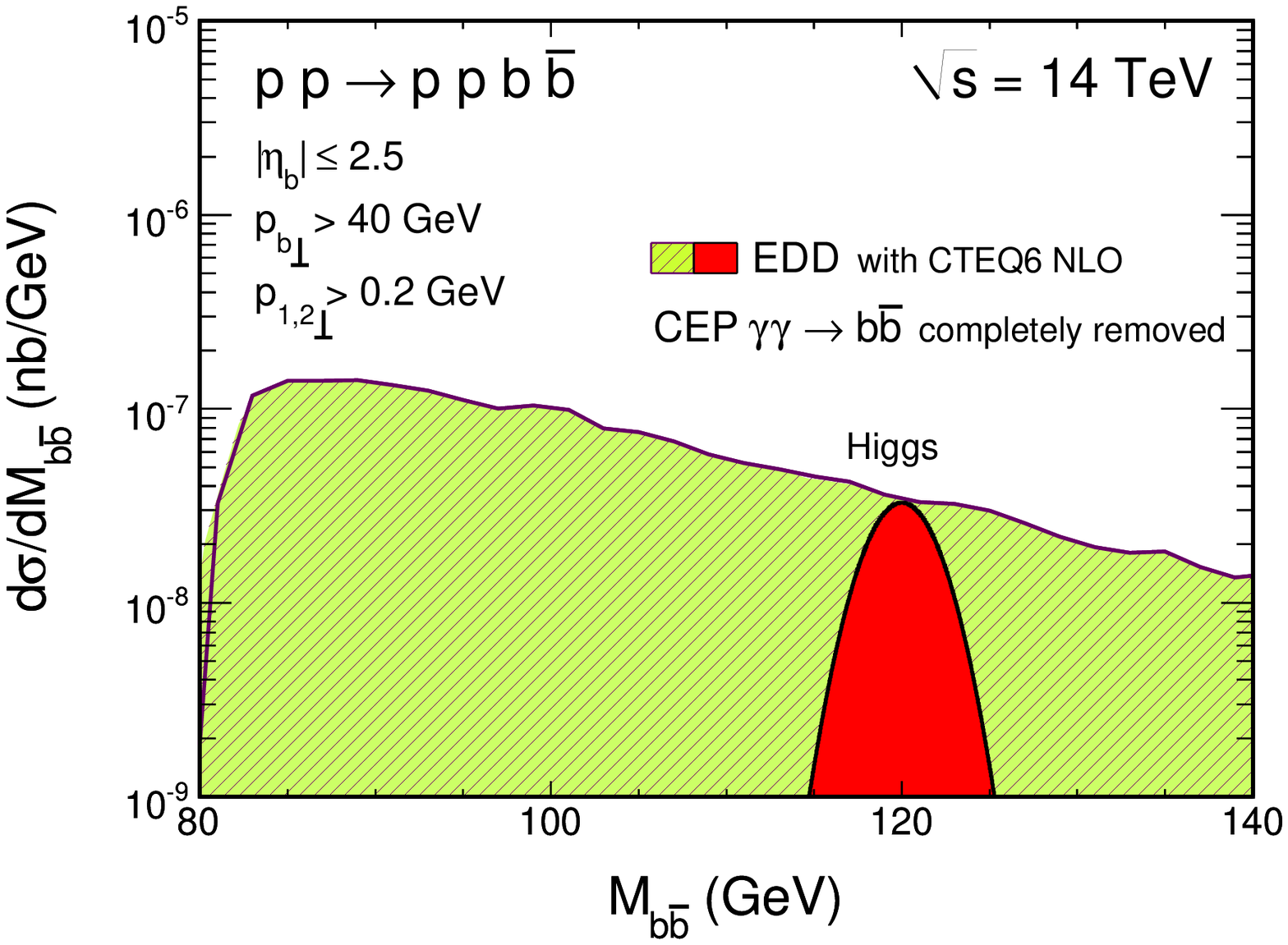}
\end{center}
\caption{Invariant mass distribution of the $b \bar b$ dijets.
Here experimental resolution on $p p$ missing mass has been included.
The Gaussian-type line (area) corresponds to the Standard Model Higgs boson 
production.
Shown are results with different combinations of cuts which can be used to
improve the signal-to-background ratio.}
\label{fig:dsig_dMbb_cuts}
\end{figure}

In Refs.\cite{MPS10higgs,MPS11higgs} we have discussed in great detail
how to improve the difficult situation. Examples are shown in 
Fig.\ref{fig:dsig_dMbb_cuts}. 
In all considered cases the situation seems much better. We have checked,
however, that this is an optimal situation and further imporovement
of the signal-to-background ratio is not possible.

\subsection{Digluon production}

\begin{figure} [!thb]
\begin{center}
\includegraphics[width=5.5cm]{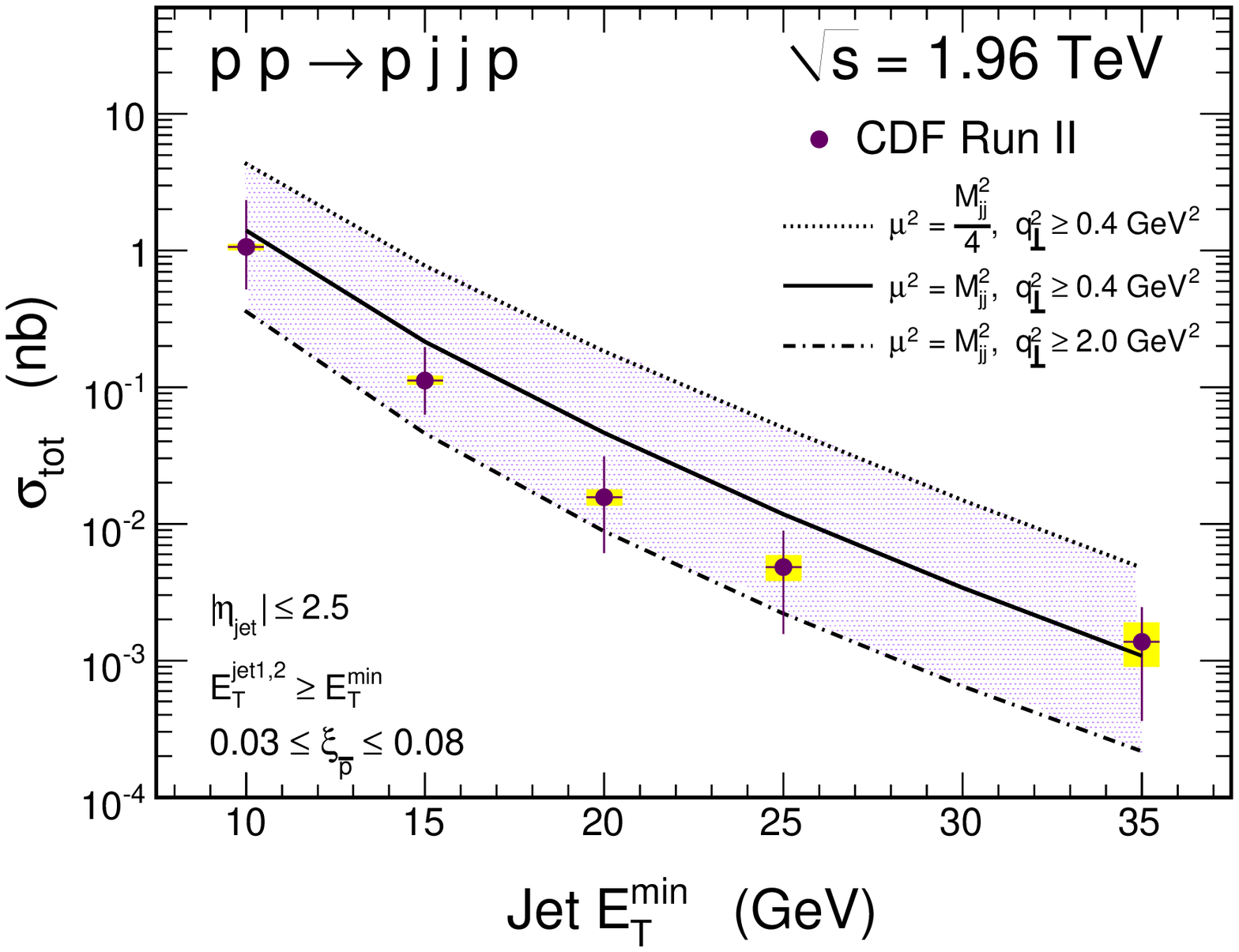} 
\includegraphics[width=5.5cm]{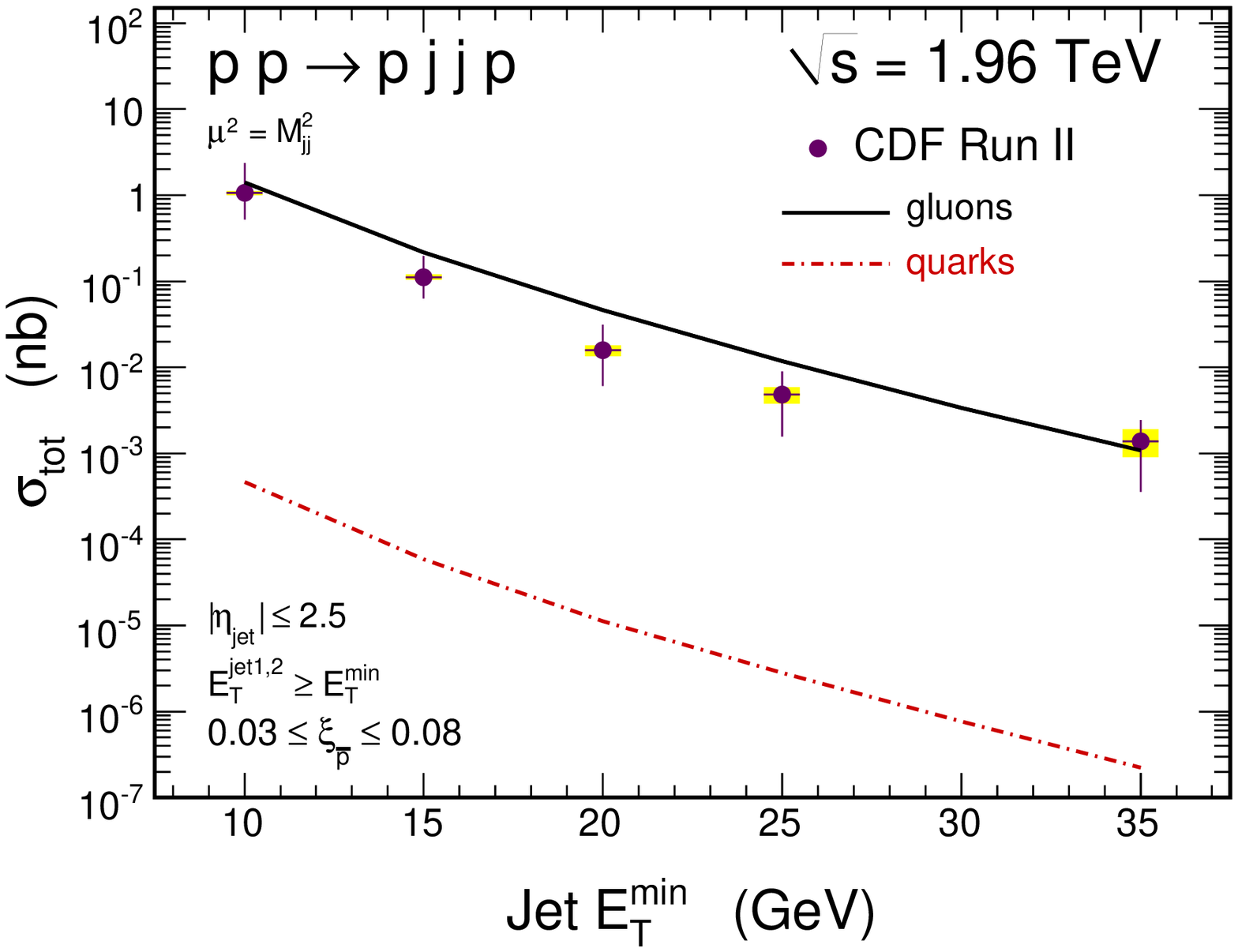} 
\end{center}
\caption{Cross section as a function of the lower cut on jet transverse
energy. The left panel shows uncertainties on the choice of scales while
the right panel compares the digluon and quark-antiquark components.}
\label{fig:sig_ETlower}
\end{figure}

The dijet production has been measured by the CDF collaboration
at the Tevatron \cite{CDF-dijets}. In Fig.\ref{fig:sig_ETlower}
we show the experimental data together with our predictions.
We get a good description of the data within model uncertainties.
We find that the quark-antiquark contribution is more than three orders
of magnitude smaller than the digluon one.


\begin{figure}
\begin{center}
\includegraphics[width=6cm]{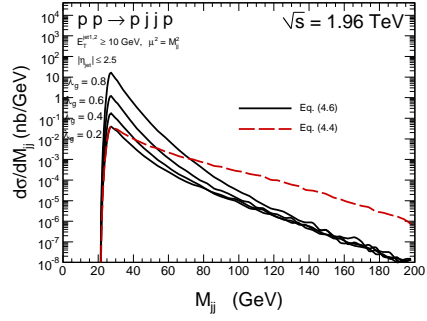} 
\end{center}
\caption{Invariant mass distribution of gluonic dijets.}
\label{fig:dsig_dMgg}
\end{figure}

In Fig.\ref{fig:dsig_dMgg} we present invariant mass distribution.
We compare contribution of the two mechanisms shown in Fig.\ref{fig:gg_CEP}.
The contribution of the diagram B depends strongly on the power of
(collinear) gluon distribution at low-$x$. It is clear that its
contribution at larger invariant masses is completely negligible.


\begin{figure}[h!]
\begin{center}
\includegraphics[width=6.0cm]{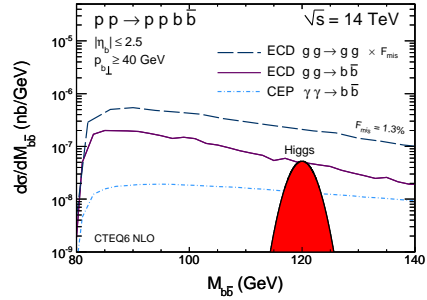}
\end{center}
 \caption{
Invariant mass distribution of the $b \bar b$ system. Shown are
contributions from diffractive Higgs boson (shaded area), $b \bar b$
continuum (solid line), $\gamma \gamma$ continuum (dash-dotted line) and
diffractive digluon contribution (dashed line) multiplied by an ATLAS
misidentification factor squared.
}
\label{fig:dsig_dMjj_Higgs}
\end{figure}

In Ref.\cite{MPS11_gluons} we have discussed also a formally reducible
background of the gluonic dijets to the Higgs boson production.
This contribution is similar to that for the $b \bar b$ continuum
discussed also in this presentation above. 
In Fig.\ref{fig:dsig_dMjj_Higgs} we show an example of the reducible 
background with some set of cuts.
Large contribution of the reducible background has been found.

\section{Conclusions}

We have shown and discussed differential distributions for the continuum
$b \bar b$ production. The corresponding amplitude has been calculated
in the Khoze-Martin-Ryskin approach.

The $b \bar b$ continuum constitutes irreducible background for
exclusive Higgs boson production.
Experimental resolution on $ pp$ missing mass has been included when 
comparing the Higgs signal and the $b \bar b$ background.
Our analysis shows that a special cuts can be useful to see the Standard
Model Higgs boson signal. 


We have discussed also production of gluonic dijets. In our approach 
to corresponding matrix element has been calculated using rather Lipatov 
reggeon-reggeon-gluon vertices. We have considered a new mechanism when
gluons are emitted from different $t$-channel gluons (reggeons).
The contribution of the latter mechanism turned out to be rather small.

When gluons are missidentified as $b$ ($\bar b$) jets the exclusive
digluon continuum constitutes a reducible background to exclusive 
Higgs boson production. We have shown that this background is comparable
to the irreducible $b \bar b$ continuum. 

Our analysis indicates that a real experiment can be rather difficult.
The situation could be better for some scenarios beyond the
Standard Model \cite{Cox:2007sw,HKRSTW08}. 

\vspace{0.5cm}

I am indebted to Rafa{\l} Maciu{\l}a and Roman Pasechnik
for collaboration on the issues presented here.
I congratulate the organizers, Tran Thanh Vahn, Chung-I Tan and Mary Rotondo,
for very good organization and friendly atmosphere during the Blois
workshop in Quy Nhon.


\end{document}